\documentclass[12pt,preprint]{aastex}
\def\ergcm2s{${\rm\,ergs\,cm^{-2}\,s^{-1}}$}

\newcommand{\obj}{IRAS F01475-0740}

\begin{document}

\title{A unique X-ray unabsorbed Seyfert 2 galaxy IRAS F01475-0740}

\author{Huang, Xing-Xing\altaffilmark{1}, Wang, Jun-Xian\altaffilmark{1}, Tan, Ying\altaffilmark{1},  Yang, Huan\altaffilmark{1}, and Huang, Ya-Fang\altaffilmark{2}}
\altaffiltext{1}{CAS Key Laboratory for Research in Galaxies and Cosmology, Department of Astronomy,
University of Science and Technology of China, Hefei, Anhui 230026, China; hxx@mail.ustc.edu.cn, jxw@ustc.edu.cn.}
\altaffiltext{2}{National Astronomical Observatories, Chinese Academy of Sciences, A20 Datun Road, Beijing 100012, China}

\begin{abstract}
X-ray unabsorbed Seyfert 2 galaxies appear to have X-ray absorption
column densities that are too low ($N_{\rm H}<10^{22}$ cm$^{-2}$) to
explain the absence of broad emission lines in their optical spectra,
challenging the standard AGN unification model.
In this paper we report Suzaku exposure on the X-ray unabsorbed Seyfert 2 galaxy \obj, in which a hidden broad line region was detected through spectropolarimetric observation.
The X-ray data show rapid and significant variations on time scales
down to 5 ks, indicating that we are viewing its central engine directly.
A newly obtained optical spectrum and previous optical/X-ray
data suggest that state transition is unlikely in this source.
These make \obj\ a very peculiar X-ray unabsorbed Seyfert 2 galaxy which can only be explained by absorption from materials with abnormally high dust-to-gas ratio (by a factor of $>$ 4 larger than Galactic).
 This is in contrast to most AGNs, which typically show dust-to-gas ratios 3
-- 100 times lower than the Galactic.
 \end{abstract}


\section{Introduction}
The Active Galactic Nucleus (AGN) unification scenario has been successful for explaining different types of active galaxies \citep{Antonucci93}. In the unification model, type 1 and type 2 AGNs are believed intrinsically the same  but viewed from different orientations, and the absence of broad emission lines (BELs) in type 2 AGNs is due to the  obscuration by an optically thick structure (dusty torus) from our line of sight.
The most convincing evidence supporting the unified model is the detection of polarized BELs in Seyfert 2 galaxies by spectropolarimetric observations \citep[e.g.][]{Antonucci85, Tran01}.
X-ray observations also support the unified scenario by detecting typically larger absorption column densities $N_{H}>10^{23}\,cm^{-2}$ in type 2 AGNs \citep[e.g.][]{Risaliti99}.

However, a fraction of Seyfert 2 AGNs were reported to show no or very low X-ray absorption
(with $N_{H}<10^{22}\,cm^{-2}$) \citep{panessa02, Hawkins04, Brightman08, Tran10}, apparently in contrast to
the standard unification model.
Possible explanations to these ``X-ray unabsorbed Seyfert 2" include:

1) They appear ``unabsorbed" in X-ray due to contamination from the
host galaxy, or a scattering component \citep[e.g.][]{shu10}.

2) Weak/absent BLRs in low luminosity or low accretion rate sources \citep{Nicastro00}.

3) State transitions and non-simultaneous X-ray and optical
observations, i.e., the absence of broad emission lines could be due
to a very low state in the central engine \citep[e.g.][]{Gilli00}.
 X-ray absorption was also found variable rapidly in some sources \citep[e.g.][]{Maiolino10}.

4) Extremely high dust to gas ratio  (or expressed as E$_{B-V}$/N$_H$ or A$_V$/N$_H$) compared with the Galactic value. However this is in contradiction with observations that \citet{Maiolino01} measured E$_{B-V}$/N$_H$ of 3 -- 100 times lower than Galactic for
obscured AGNs at intermediate and high luminosities.

{\obj} was identified as a Seyfert 2 galaxy at $z=0.017666$ by \citet{Aguero96} with the 2.15-m CASLEO
telescope in Argentina.
\citet{Tran01} reported the detection of polarized broad emission lines in {\obj}, indicating the existence of a hidden BLR.
\obj\ was identified as X-ray unabsorbed based on XMM-Newton observation.
\citet{Gu05} announced that the X-ray spectra observed by XMM-Newton in 2004
showed low X-ray absorption ($N_{\rm H}=4.7^{+0.5}_{-0.4}\times10^{21} cm^{-2}$).
As noted by Brightman \& Nandra (2008), such low X-ray absorption
is insufficient to suppress the optical broad lines unless unless the absorber has an anomalously high dust-to-gas ratio.
They instead proposed that \obj\ is heavily obscured and its
X-ray ``unabsorbed" appearance is due to a scattering component by the electrons which also scatter broad optical emission lines.

In this paper we present a new Suzaku observation on \obj\ taken in 2008, in which we detect significant and rapid X-ray variations at time scale down to 5 ks. Such rapid variations, which were not seen during previous XMM exposure, provide strong constraint
to the X-ray absorption nature of \obj.

\section{Suzaku Data Reduction}

\obj\ was observed with Suzaku on July 14th, 2008 for a net XIS exposure time of 57.9 ks (ObsID 703065010).
 We reprocessed the XIS 0,1,3 data from the unfiltered event files\footnote{
 There was no useful data from XIS2 after 2006 November because of a charge leak that
occurred, see ftp://legacy.gsfc.nasa.gov/suzaku/nra\_info/suzaku\_td.pdf}.
In this letter we focus on XIS data only since \obj\ was non-detected by HXD-PIN.
All the data reduction followed the standard procedure
illustrated in the Suzaku reduction guide\footnote{http://heasarc.gsfc.nasa.gov/docs/suzaku/analysis/abc/}.
The most recent calibration released on 2010 November 04 and
the analysis software package HEASOFT 6.10 were used.

We examined the XIS 0.4--2 keV and 2--10 keV images from cleaned event files and
find no contamination from nearby sources. This is also confirmed in the XMM image.
To extract XIS spectra, we combined the $3\times3$ and $5\times5$ modes for each CCD.
The source spectra were extracted from circular regions with radius of 3.0\arcmin,
and the background from larger circular regions
excluding the source and avoiding the calibration sources in corners.
The response and ancillary response files were created finally
using $xisrmfgen$ and $xissimarfgen$.

\section{Spectral Fitting and Rapid X-ray Variation}

In this paper the spectra of XIS0 and XIS3 (both are front-illuminated) were always added with $addspec$, and fitted simultaneously with the spectrum of XIS1 (back-illuminated).
In the spectral fitting, we chosed 0.4-10 keV band for XIS0+XIS3 spectrum,  and 0.4-8 keV band for XIS1 (due to its much lower effective area at higher energy).
The 1.5 -- 2.0 keV spectra were excluded because of calibration uncertainty due to the Si K edge \citep{Koyama07}.
Fitting errors are of $90\%$ confidence level for one parameter ($\Delta{\chi}^2=2.71$).
The adopted cosmological parameters are $H_0=70 \ km\ s^{-1} Mpc^{-1}$, $\Omega_\Lambda=0.73$ and $\Omega_m=0.27$.
During spectral fitting, the Galactic column density along the line of sight to {\obj}
is always taken into account \citep[$N_{\rm H}=2.32\times10^{20}\ cm^{-2}$,][]{Dickey90}.

The 0.4--10 keV XIS 
spectra integrated over the whole observation are shown in Fig. 1.
The spectra were first fitted with an absorbed  power-law with
$\Gamma=2.23^{+0.08}_{-0.08}$ and $N_H=73^{+6}_{-6}\times10^{20}\,cm^{-2}$.
An additional soft component ($mekal$ in XSPEC with $kT=0.30^{+0.08}_{-0.08}$ keV) is statistically required. Such a soft component could be due to emission from the
host galaxy, or due to a soft excess from the AGN \citep[also see][]{Teng10}.
Narrow iron $K_\alpha$ line was non-detected, and we provide the upper limit in Table 1 to its equivalent width  by fixing the line width $\sigma$ at 0.02 keV and central energy at 6.4 keV.

The Suzaku XIS light curves in the 0.5--2 keV and 2--10 keV bands are plotted
in Fig. 2, in which we see significant variations in both bands and in all three XIS detectors\footnote{Fitting each light curve with a constant yields large $\chi^2$ $>$ 230 for 19 data bins},
on time scale down to 5 ks.
The amplitude of the variation is much larger than the background count rate, thus the rapid variation can not be due to the variation in
instrumental background.

To investigate for spectral variation, we extract X-ray spectra in three time intervals signed High/Middle/Low state in Fig. 2.
We fit them separately with the $mekal$ component fixed to the best-fit model of the whole exposure, and the fitting results are given in Table 1.
{\obj} was also observed by XMM-newton in January 2004.
We reprocessed the spectra of PN and MOS as \citet{Gu05} and the fitting results with the same model are also included in Table 1. We note that XMM fluxes are consistent with Suzaku low state.
We saw no significant variation in spectral shape (photon index or absorption) within Suzaku exposure or between Suzaku and XMM observations.
This indicates that the rapid X-ray variation is intrinsic, and can not be attributed to variation
in absorption, as detected in several other sources \citep[e.g.][]{Risaliti10}.
Note that a rapid X-ray variation in \obj\ was also detected during ROSAT PSPC exposure on July 13th, 1992.
\citet{Pfefferkorn01} found its ROSAT count rate
in the 0.2--2 keV band decreased from 0.064 to 0.021 counts/s within 12.9 hours, with the amplitude of the variation similar to Suzaku data.

\section{Discussion}

We first note that the rapid X-ray variation in \obj\ is unlikely due to contamination from a nearby X-ray source, which must have comparable but variable X-ray flux, and locate close to \obj\ by coincidence. The ROSAT PSPC image obtained in 1992 has better resolution (25$\arcsec$ at 1 keV) than Suzaku (1.6$\arcmin$). On PSPC image we clearly see a single point  source at the position of \obj, without contamination from nearby sources within 4$\arcmin$.
The XMM MOS images obtained in 2004 have the best spatial resolution (6$\arcsec$ FWHM) in X-ray available for \obj. Although XMM data revealed no variation \citep{Brightman08}, no nearby sources were detected within
4$\arcmin$, and \obj\ remained point-like during the exposure.
These indicate that, if there is any nearby source which is responsible for the detected rapid variation during ROSAT and Suzaku observations,  either its distance to \obj\ is too small ($<$ 6$\arcsec$) to be resolved by XMM MOS, or it could locate at a larger distance  ($<$ 25\arcsec, still unresolvable with ROSAT PSPC though) but totally disappear during XMM exposure.
The X-ray source number density with f$_{2-10keV}$ above 1 $\times$ 10$^{-12}$ \ergcm2s\ provided by ASCA Large Sky Survey is 0.5 deg$^{-2}$
\citep{Ueda99}. This means that the possibility to have another X-ray bright source within 25\arcsec\ to \obj\ by chance is 0.00015. The possibility further drops to 9 $\times$ 10$^{-6}$ for a distance of 6\arcsec.  We also note that the observed luminosity L$_{2-10keV}$ varies from 1.7 to 0.5 $\times$ 10$^{42}$ erg s$^{-1}$ during Suzaku exposure. Such luminosity is too high for an off-nuclear Ultra Luminous X-ray source (ULX) located in \obj\ \citep{Liu05, Liu06}.

\citet{Wang07} estimated a mass of 10$^{7.55}\,M_\odot$ for the super-massive black hole (SMBH) in the nucleus of {\obj}  (deduced from the mass of bulge). For such a SMBH, the observed time scale of rapid X-ray variation (5 ks) corresponds to $\sim$ 70 r$_g$ (r$_g$ = 2GM/c$^2$). This is  far smaller than the sizes of the BLR, the torus, and the region of scattering electrons.

The rapid and significant X-ray variation detected in \obj\ thus proves that we were viewing its central engine directly (without significant
contribution to X-ray flux from the host galaxy), and the fitted small X-ray absorption column density is physical instead of polluted by emission from host galaxy or scattering/reflection component \citep[e.g.][]{shu10, Brightman08}.
The small iron $K_\alpha$ EW and the X-ray to reddening-corrected [OIII] flux ratio ($f_{2-10\,keV}/f_{[OIII]}$ = 1 -- 3)
are consistent with (e.g. do not oppose) its Compton-thin identity \citep{Bassani99}.
Note that based on the $f_{2-10keV}/f_{IR}$ versus $f_{[OIII]}/f_{IR}$ plot, \cite{Petrov09} also classified \obj\ as Compton-thin with X-ray emission dominated by AGN process.

In \S1 we have listed possible explanations for ``X-ray unabsorbed S2" galaxies.
The first of them, i.e., due to contamination from host galaxy or scattering/reflection component in X-ray, can thus be ruled out based the detected rapid X-ray variation. The second one, i.e., intrinsically weak/lack of BLR in low luminosity or low accretion rate sources is not applicable either.
This is because that  polarized broad emission line has been detected \citep{Tran01}.
We further note that the luminosity and accretion rate of \obj\ (with L$_{2-10keV}$ = 0.5 -- 1.7 $\times$ 10$^{42}$ erg s$^{-1}$, L$_{[OIII]}$ = 5 $\times$ 10$^{41}$ erg s$^{-1}$,
and an Eddington ratio of 0.28)\footnote{An Eddington ratio of 0.28 is estimated based on [OIII] emission (see Wang \& Zhang 2007). Assuming a 10-200 correction factor from 2-10 keV to bolometric luminosity \citep{Lusso10}, the Eddington ratio for this object is  $~0.4\%-8\%$.}
are consistent with those type 2 AGNs with hidden BLR detected \citep{Shu07, Nicastro03}.
 Below we discuss the rest two possibilities.

\subsection{State Transition?}

In this sub-section we examine whether there were state transitions in \obj.
We obtained a new optical spectrum with the NAOC (National Astronomical Observatories, Chinese Academy of Sciences) 2.16m telescope on 21st July, 2010.
The spectrum with a resolution of 9.6 {\AA} is plotted in Fig. 4, in which no broad components of the permit lines were detected.

The detected narrow Balmer decrement is consistent with \citet{Grijp92} (the number in brackets is from de Grijp et al.):
$\frac{H_\alpha}{H_\beta}=7.21$ (7.19).
We convert the Balmer decrement to optical reddening using
the formula of \citet{Ward87}: $A_V=6.67\times(log(H_\alpha/H_\beta)-log(2.85))\,mag$,
and obtained an optical extinction $A_V=2.7$ for the narrow line region.
The reddening to the narrow line region corresponds to an X-ray absorption column density of $N_H\sim6\times10^{21} cm^{-2}$
 \citep[assuming a Galactic dust to gas ratio with $A_V= 4.5\pm0.3 \times10^{-22}N_H\,cm^{-2}$,][]{Gorenstein75}.
This indicates that the observed X-ray absorption is only sufficient for the reddening to NLR assuming a Galactic dust to gas ratio. The lack of broad emission lines in the optical spectra obviously requires much higher reddening.

\citet{Rush96} reported ROSAT PSPC spectrum of \obj\ obtained in July 1992 with a total integrated exposure time of 6.9 ks.
The PSPC  flux in 0.1 -- 2 keV is 4.6 $\times$ $10^{-13}$\ergcm2s, similar to the
soft X-ray flux detected by Suzaku (see Table 1).
We conclude that based on the optical spectra (one in March 1993 and one in July 2010),  and the consistent X-ray unabsorbed appearance and/or fluxes detected by ROSAT (July 1992), XMM (Jan. 2004) and Suzaku (July 2008) respectively, the state transition scenario is unlikely for \obj. Simultaneously X-ray and optical data are required though to completely rule out this scenario.

Alternatively, the absorber in \obj\ could be clumpy, and the low X-ray
N$_H$ could be due to lower  column density holes in the absorber. However, this is unlikely either since while the non-detection of broad emission lines in optical spectra needs almost complete coverage of the absorber to the BLR, consistent low N$_H$ detected by multiple X-ray observations requires the absorber to be quite clumpy (that we have higher chance to view the central engine in X-ray through low column density
holes). Spectral fitting to Suzaku data also indicates that partial covering absorption is statistically not required ($\Delta\chi^2$ $<$ 1).

\subsection{Abnormally high dust-to-gas ratio}
The only possibility left is that the absorption material in \obj\ has abnormally high dust-to-gas ratio. Based on the new optical spectrum we obtained with NAOC 2.16m telescope, we derived an upper limit to the broad H$\alpha$ to narrow [NII] 6583 line flux ratio of 0.14 (assuming the broad H$\alpha$ has a FWHM of 5000 km/s). This upper limit is 180 time lower than the typical value in Seyfert 1 galaxies \citep{Osterbrock77}. This requires an extra extinction of A$_V$ $>$ 6.9 (assuming a Galactic extinction curve) to the hidden broad line region, in additional to $A_V=2.7$ to the narrow line region. Consequently the dust to gas ratio in \obj\ is at least $\sim$4 times larger than Galactic.

However, such high dust-to-gas ratio is rather unusual among AGNs.
\citet{Maiolino01} report that while E$_{B-V}$/N$_H$ in low luminosity AGNs (with L$_X$ $\sim$ 10$^{41}$ erg s$^{-1}$) could be several times higher than the Galactic, AGNs with
 L$_X$ $>$ 10$^{42}$ erg s$^{-1}$ have E$_{B-V}$/N$_H$ lower than Galactic by a factor of 3 -- 100. Note that absorption in AGNs with dust-to-gas ratio consistent with Galactic were also reported (e.g. Wang et al. 2009).
 It is interesting to note that the  X-ray luminosity of {\obj} (L$_{2-10\,keV}=0.5\,-\,1.7 \times 10^{42}$ erg s$^{-1}$) is right on the border between low luminosity sources with E$_{B-V}$/N$_H$ higher than Galactic and high luminosity sources with E$_{B-V}$/N$_H$ lower than Galactic in \citet{Maiolino01}.

 \citet{Brandt96} proposed that dusty warm absorber could exist in the quasar IRAS 13349+2438, to explain the lack of X-ray cold absorption in contradiction with strong optical reddening. Several similar cases were later reported \citep{Komossa99}. To examine whether the absorber in \obj\ is also ionized, we fit the Suzaku spectra with an ionized absorption model $warmabs$. We  find that warm absorption is statistically not required. X-ray absorption edges due to ionized gas or dust were non-detected either in Suzaku or XMM spectra.

We finally note that \obj\ also appears peculiar in infrared spectrum. It is the only Seyfert 2 galaxy with mid-IR silicate emission in the Spitzer sample of \citet{Hao07}. This makes it more like a type 1 AGN, suggesting smaller dust absorption opacity in IR comparing with other Seyfert 2 galaxies. This is however consistent with the small X-ray absorption detected, suggesting that the dust might be dominated by small grains which could steepen the optical/IR extinction curve.
Broad emission lines are expected to be detectable in near-IR spectra for absorber with small column density \citep{Lutz02}. Deep near-IR and optical spectra are essential to  detect the reddened broad emission line and measure the abnormal dust extinction.

\acknowledgments
The work was supported by Chinese NSF through Grant 10773010/10825312,
and the Knowledge Innovation Program of CAS (Grant No. KJCX2-YW-T05). We thank Dr. Hai Fu for the help to improve the manuscript.


\begin{thebibliography}{38}
\expandafter\ifx\csname natexlab\endcsname\relax\def\natexlab#1{#1}\fi

\bibitem[{{Aguero} {et~al.}(1996){Aguero}, {Paolantonio}, \&
  {Suarez}}]{Aguero96}
{Aguero}, E.~L., {Paolantonio}, S., \& {Suarez}, F. 1996, \pasp, 108, 1117

\bibitem[{{Antonucci}(1993)}]{Antonucci93}
{Antonucci}, R. 1993, \araa, 31, 473

\bibitem[{{Antonucci} \& {Olszewski}(1985)}]{Antonucci85}
{Antonucci}, R.~R.~J., \& {Olszewski}, E.~W. 1985, \aj, 90, 2203

\bibitem[{{Bassani} {et~al.}(1999){Bassani}, {Dadina}, {Maiolino}, {Salvati},
  {Risaliti}, {della Ceca}, {Matt}, \& {Zamorani}}]{Bassani99}
{Bassani}, L., {Dadina}, M., {Maiolino}, R., {Salvati}, M., {Risaliti}, G.,
  {della Ceca}, R., {Matt}, G., \& {Zamorani}, G. 1999, \apjs, 121, 473

\bibitem[{{Brandt} {et~al.}(1996){Brandt}, {Fabian}, \& {Pounds}}]{Brandt96}
{Brandt}, W.~N., {Fabian}, A.~C., \& {Pounds}, K.~A. 1996, \mnras, 278, 326

\bibitem[{{Brightman} \& {Nandra}(2008)}]{Brightman08}
{Brightman}, M., \& {Nandra}, K. 2008, \mnras, 390, 1241

\bibitem[{{de Grijp} {et~al.}(1992){de Grijp}, {Keel}, {Miley}, {Goudfrooij},
  \& {Lub}}]{Grijp92}
{de Grijp}, M.~H.~K., {Keel}, W.~C., {Miley}, G.~K., {Goudfrooij}, P., \&
  {Lub}, J. 1992, \aaps, 96, 389

\bibitem[{{Dickey} \& {Lockman}(1990)}]{Dickey90}
{Dickey}, J.~M., \& {Lockman}, F.~J. 1990, \araa, 28, 215

\bibitem[{{Gilli} {et~al.}(2000){Gilli}, {Maiolino}, {Marconi}, {Risaliti},
  {Dadina}, {Weaver}, \& {Colbert}}]{Gilli00}
{Gilli}, R., {Maiolino}, R., {Marconi}, A., {Risaliti}, G., {Dadina}, M.,
  {Weaver}, K.~A., \& {Colbert}, E.~J.~M. 2000, \aap, 355, 485

\bibitem[{{Gorenstein}(1975)}]{Gorenstein75}
{Gorenstein}, P. 1975, \apj, 198, 95

\bibitem[{{Guainazzi} {et~al.}(2005){Guainazzi}, {Matt}, \& {Perola}}]{Gu05}
{Guainazzi}, M., {Matt}, G., \& {Perola}, G.~C. 2005, \aap, 444, 119

\bibitem[{{Hao} {et~al.}(2007){Hao}, {Weedman}, {Spoon}, {Marshall},
  {Levenson}, {Elitzur}, \& {Houck}}]{Hao07}
{Hao}, L., {Weedman}, D.~W., {Spoon}, H.~W.~W., {Marshall}, J.~A., {Levenson},
  N.~A., {Elitzur}, M., \& {Houck}, J.~R. 2007, \apjl, 655, L77

\bibitem[{{Hawkins}(2004)}]{Hawkins04}
{Hawkins}, M.~R.~S. 2004, \aap, 424, 519

\bibitem[{{Komossa}(1999)}]{Komossa99}
{Komossa}, S. 1999, in Astronomical Society of the Pacific Conference Series,
  Vol. 175, Structure and Kinematics of Quasar Broad Line Regions, ed.
  {C.~M.~Gaskell, W.~N.~Brandt, M.~Dietrich, D.~Dultzin-Hacyan, \&
  M.~Eracleous}, 365--+

\bibitem[{{Koyama} {et~al.}(2007){Koyama}, {Tsunemi}, {Dotani}, {Bautz},
  {Hayashida}, {Tsuru}, {Matsumoto}, {Ogawara}, {Ricker}, {Doty}, {Kissel},
  {Foster}, {Nakajima}, {Yamaguchi}, {Mori}, {Sakano}, {Hamaguchi},
  {Nishiuchi}, {Miyata}, {Torii}, {Namiki}, {Katsuda}, {Matsuura}, {Miyauchi},
  {Anabuki}, {Tawa}, {Ozaki}, {Murakami}, {Maeda}, {Ichikawa}, {Prigozhin},
  {Boughan}, {Lamarr}, {Miller}, {Burke}, {Gregory}, {Pillsbury}, {Bamba},
  {Hiraga}, {Senda}, {Katayama}, {Kitamoto}, {Tsujimoto}, {Kohmura}, {Tsuboi},
  \& {Awaki}}]{Koyama07}
{Koyama}, K., {et~al.} 2007, \pasj, 59, 23

\bibitem[{{Liu} {et~al.}(2006){Liu}, {Bregman}, \& {Irwin}}]{Liu06}
{Liu}, J., {Bregman}, J.~N., \& {Irwin}, J. 2006, \apj, 642, 171

\bibitem[{{Liu} \& {Mirabel}(2005)}]{Liu05}
{Liu}, Q.~Z., \& {Mirabel}, I.~F. 2005, \aap, 429, 1125

\bibitem[{{Lusso} {et~al.}(2010){Lusso}, {Comastri}, {Vignali}, {Zamorani},
  {Brusa}, {Gilli}, {Iwasawa}, {Salvato}, {Civano}, {Elvis}, {Merloni},
  {Bongiorno}, {Trump}, {Koekemoer}, {Schinnerer}, {Le Floc'h}, {Cappelluti},
  {Jahnke}, {Sargent}, {Silverman}, {Mainieri}, {Fiore}, {Bolzonella}, {Le
  F{\`e}vre}, {Garilli}, {Iovino}, {Kneib}, {Lamareille}, {Lilly}, {Mignoli},
  {Scodeggio}, \& {Vergani}}]{Lusso10}
{Lusso}, E., {et~al.} 2010, \aap, 512, A34+

\bibitem[{{Lutz} {et~al.}(2002){Lutz}, {Maiolino}, {Moorwood}, {Netzer},
  {Wagner}, {Sturm}, \& {Genzel}}]{Lutz02}
{Lutz}, D., {Maiolino}, R., {Moorwood}, A.~F.~M., {Netzer}, H., {Wagner},
  S.~J., {Sturm}, E., \& {Genzel}, R. 2002, \aap, 396, 439

\bibitem[{{Maiolino} {et~al.}(2001){Maiolino}, {Marconi}, {Salvati},
  {Risaliti}, {Severgnini}, {Oliva}, {La Franca}, \& {Vanzi}}]{Maiolino01}
{Maiolino}, R., {Marconi}, A., {Salvati}, M., {Risaliti}, G., {Severgnini}, P.,
  {Oliva}, E., {La Franca}, F., \& {Vanzi}, L. 2001, \aap, 365, 28

\bibitem[{{Maiolino} {et~al.}(2010){Maiolino}, {Risaliti}, {Salvati},
  {Pietrini}, {Torricelli-Ciamponi}, {Elvis}, {Fabbiano}, {Braito}, \&
  {Reeves}}]{Maiolino10}
{Maiolino}, R., {et~al.} 2010, \aap, 517, A47+

\bibitem[{{Nicastro} \& {Elvis}(2000)}]{Nicastro00}
{Nicastro}, F., \& {Elvis}, M. 2000, \nar, 44, 569

\bibitem[{{Nicastro} {et~al.}(2003){Nicastro}, {Martocchia}, \&
  {Matt}}]{Nicastro03}
{Nicastro}, F., {Martocchia}, A., \& {Matt}, G. 2003, \apjl, 589, L13

\bibitem[{{Osterbrock}(1977)}]{Osterbrock77}
{Osterbrock}, D.~E. 1977, \apj, 215, 733

\bibitem[{{Panessa} \& {Bassani}(2002)}]{panessa02}
{Panessa}, F., \& {Bassani}, L. 2002, \aap, 394, 435

\bibitem[{{Petrov}(2009)}]{Petrov09}
{Petrov}, G.~P. 2009, Bulgarian Astronomical Journal, 11, 79

\bibitem[{{Pfefferkorn} {et~al.}(2001){Pfefferkorn}, {Boller}, \&
  {Rafanelli}}]{Pfefferkorn01}
{Pfefferkorn}, F., {Boller}, T., \& {Rafanelli}, P. 2001, \aap, 368, 797

\bibitem[{{Risaliti}(2010)}]{Risaliti10}
{Risaliti}, G. 2010, in American Institute of Physics Conference Series, Vol.
  1248, American Institute of Physics Conference Series, ed. {A.~Comastri,
  L.~Angelini, \& M.~Cappi}, 351--354

\bibitem[{{Risaliti} {et~al.}(1999){Risaliti}, {Maiolino}, \&
  {Salvati}}]{Risaliti99}
{Risaliti}, G., {Maiolino}, R., \& {Salvati}, M. 1999, \apj, 522, 157

\bibitem[{{Rush} \& {Malkan}(1996)}]{Rush96}
{Rush}, B., \& {Malkan}, M.~A. 1996, \apj, 456, 466

\bibitem[{{Shu} {et~al.}(2010){Shu}, {Liu}, \& {Wang}}]{shu10}
{Shu}, X.~W., {Liu}, T., \& {Wang}, J.~X. 2010, \apj, 722, 96

\bibitem[{{Shu} {et~al.}(2007){Shu}, {Wang}, {Jiang}, {Fan}, \& {Wang}}]{Shu07}
{Shu}, X.~W., {Wang}, J.~X., {Jiang}, P., {Fan}, L.~L., \& {Wang}, T.~G. 2007,
  \apj, 657, 167

\bibitem[{{Teng} \& {Veilleux}(2010)}]{Teng10}
{Teng}, S.~H., \& {Veilleux}, S. 2010, \apj, 725, 1848

\bibitem[{{Tran}(2001)}]{Tran01}
{Tran}, H.~D. 2001, \apjl, 554, L19

\bibitem[{{Tran} {et~al.}(2010){Tran}, {Lyke}, \& {Mader}}]{Tran10}
{Tran}, H.~D., {Lyke}, J.~E., \& {Mader}, J.~A. 2010, ArXiv e-prints

\bibitem[{{Ueda} {et~al.}(1999){Ueda}, {Takahashi}, {Inoue}, {Tsuru}, {Sakano},
  {Ishisaki}, {Ogasaka}, {Makishima}, {Yamada}, {Akiyama}, \& {Ohta}}]{Ueda99}
{Ueda}, Y., {et~al.} 1999, \apj, 518, 656

\bibitem[{{Wang} \& {Zhang}(2007)}]{Wang07}
{Wang}, J., \& {Zhang}, E. 2007, \apj, 660, 1072

\bibitem[{{Ward} {et~al.}(1987){Ward}, {Geballe}, {Smith}, {Wade}, \&
  {Williams}}]{Ward87}
{Ward}, M.~J., {Geballe}, T., {Smith}, M., {Wade}, R., \& {Williams}, P. 1987,
  \apj, 316, 138

\end{thebibliography}

\begin{figure}
\includegraphics[angle=-90,scale=.50]{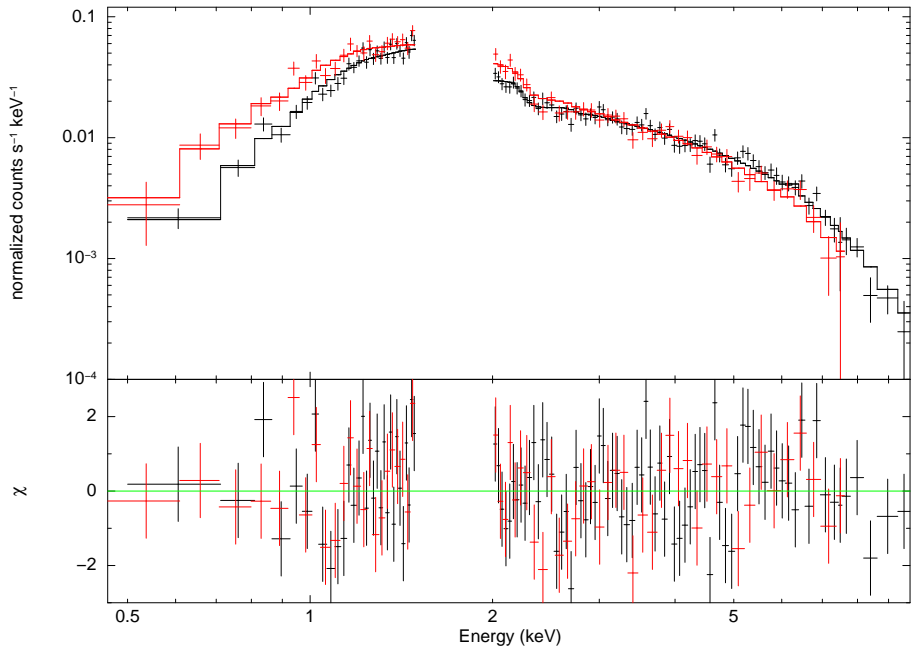}
\caption{Suzaku XIS (black for XIS0+XIS3 and red for XIS1)
spectra of \obj.
The spectra were binned to 100 counts per bin.
The data in 1.5 -- 2.0 keV were excluded because of calibration uncertainty due to the Si K edge.
The continuum model is an absorbed powerlaw plus a soft $mekal$ component.}
\end{figure}

\clearpage

\begin{figure}
\includegraphics[]{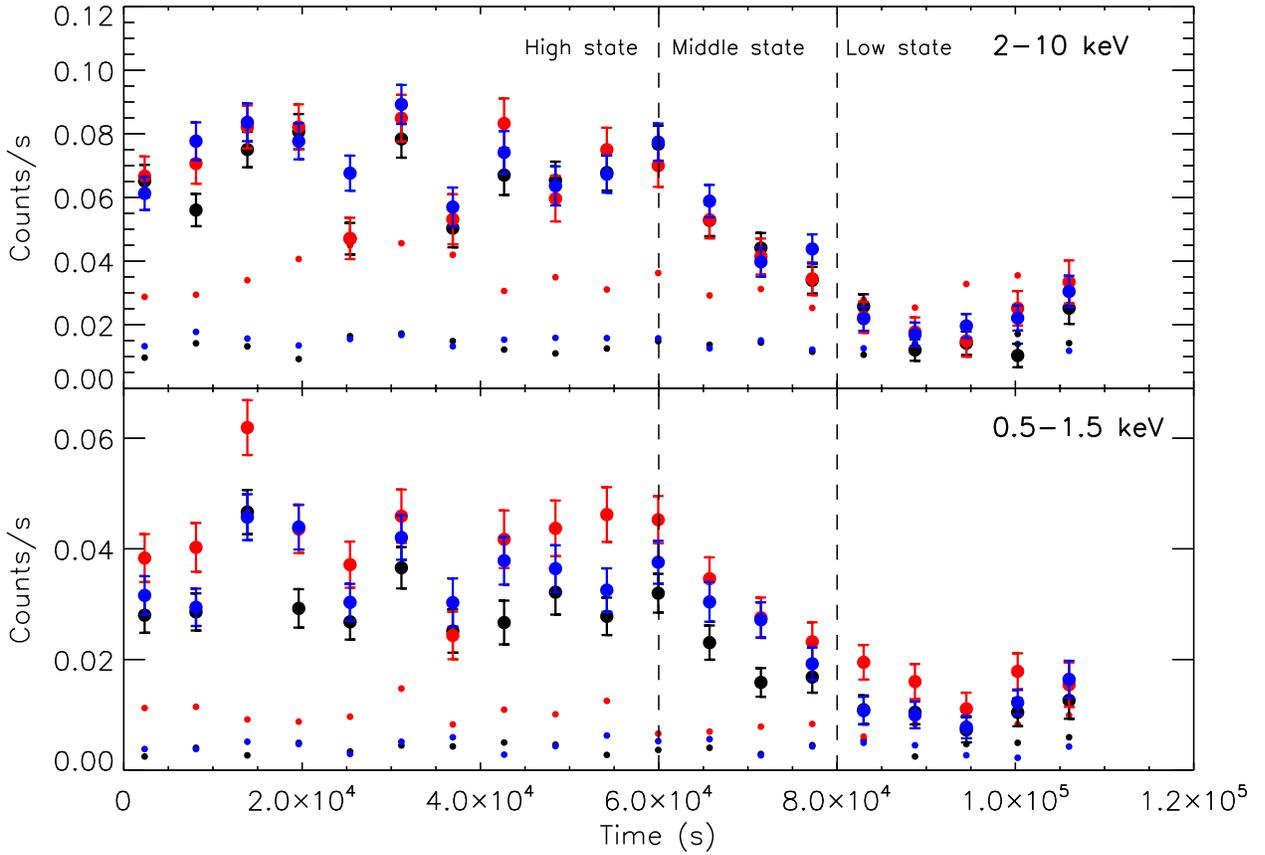}
\caption{Suzaku light curves of three XIS detectors (background subtracted,
black for XIS0, red for XIS1 and blue for XIS3) of \obj\ in the soft and hard band respectively. The bin size of the light curves is 5.76 ks which is the orbit period of Suzaku.
Light curves of expected background are over-plotted.
A rapid decline is identified at $6\times10^4$ s.  }
\end{figure}

\begin{figure}
\includegraphics[]{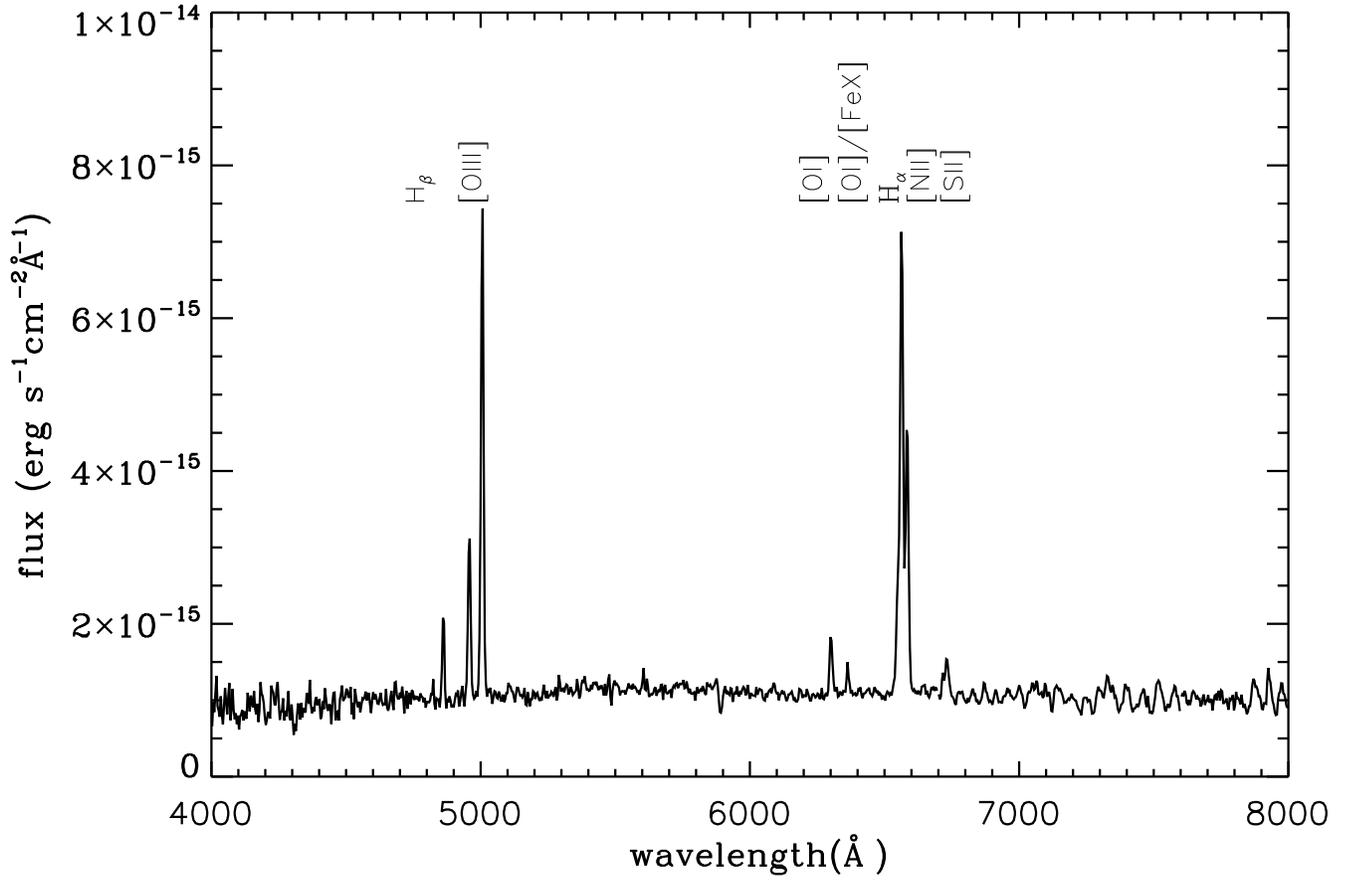}
\caption{New optical spectra of IRAS F01475-0740 obtained with the NAOC 2.16m telescope in July 2010. No broad components were detected in additional to narrow H$\alpha$ and H$\beta$ lines. The spectrum is plotted in the rest frame of \obj.}
\end{figure}

\clearpage

\begin{table}
\begin{center}
\caption{Best fit spectral parameters.\label{tbl-1}}
\begin{tabular}{crrrrrrrrrrr}
\tableline\tableline
Component &Suzaku & High State & Middle State & Low State & \multicolumn{1}{c}{XMM-Newton\tablenotemark{c}}  \\
\tableline
\multicolumn{1}{c}{$N_H$\tablenotemark{a}}        &$73^{+6}_{-6}$ &$79^{+5}_{-5}$   &$64^{+13}_{-20}$  &$41^{+11}_{-29}$ &$40^{+4}_{-4}$\\
$\Gamma$                                          &$2.23^{+0.08}_{-0.08}$ &$2.28^{+0.06}_{-0.07}$ &$2.12^{+0.20}_{-0.23}$ &$1.96^{+0.17}_{-0.36}$ &$2.04^{+0.10}_{-0.09}$\\
\multicolumn{1}{c}{F(0.5--2 keV)\tablenotemark{b}}  &$5.67^{+0.45}_{-0.60}$ &$7.50^{+0.62}_{-0.84}$   &$5.03^{+4.01}_{-1.12}$    &$2.37^{+1.59}_{-1.43}$  &$2.77^{+0.26}_{-0.33}$  \\
\multicolumn{1}{c}{F(2--10 keV)\tablenotemark{b}}   &$18.4^{+1.0}_{-1.6}$&$24.6^{+1.4}_{-1.4}$   &$16.3^{+2.7}_{-7.1}$  &$6.65^{+1.46}_{-3.73}$ &$7.70^{+0.7}_{-1.1}$ \\
$EW_{Fe K\alpha}$ (eV)$^d$     &$<64$ &$<58$ &$<510$ & $<503$ &$288^{+273}_{-267}$  \\
$kT$ (eV)                                         &$0.30^{+0.08}_{-0.08}$& 0.30$^{fixed}$ & 0.30$^{fixed}$&0.30$^{fixed}$&$0.11^{+0.15}_{-0.08}$\\
$\chi^2/dof$                                       &194/161 &108/112 &13/18&19/15 &131/152\\
\tableline
\end{tabular}
\tablenotetext{a}{Column density of the cold absorber in units of $10^{20}$ cm$^{-2}$}
\tablenotetext{b}{Flux in units of $10^{-13}$\ergcm2s.}
\tablenotetext{c}{The spectra of XMM-Newton PN/MOS are re-binned to 25 counts/bin.}
\tablenotetext{d}{With line central energy fixed at 6.4 keV, and width $\sigma$ at 0.02 keV.}
\end{center}
\end{table}

\clearpage
\end{document}